# A primitive machine learning tool for the mechanical property prediction of multiple principal element alloys

R. Tan[1], Z. Li[1], S. Zhao[1], N. Birbilis[1,2] *

[1]College of Engineering, Computing and Cybernetics, The Australian National University, Acton, ACT, 2601, Australia.

[2]Faculty of Science, Engineering, and Built Environment, Deakin University, Waurn Ponds, VIC, 3216, Australia

*nick.birbilis@deakin.edu.au

**Abstract**

Multi-principal element alloys (MPEAs) are produced by combining metallic elements in what is a diverse range of proportions. MPEAs reported to date have revealed promising performance due to their exceptional mechanical properties. Training a machine learning (ML) model on known performance data is a reasonable method to rationalise the complexity of composition dependent mechanical properties of MPEAs. This study utilises data from a specifically curated dataset, that contains information regarding six mechanical properties of MPEAs. A parser tool was introduced to convert chemical composition of alloys into the input format of the ML models, and a number of ML models were applied. Finally, Gradio was used to visualise the ML model predictions and to create a user-interactive interface. The ML model presented is an initial primitive model (as it does not factor in aspects such as MPEA production and processing route), however serves as a an initial user tool, whilst also providing a workflow for other researchers.

1. Introduction

Multi principal element alloys (MPEAs) have recently gained significant research attention, with this category of alloys also being inclusive of the so-called high-entropy alloys (HEAs) [1-6]. Whilst research into such alloy systems is entering its second decade as a widespread domain in materials and metallurgy, a number of fundamental aspects related to the composition-performance relationship in MPEAs is still emerging [7-11].

As a data science tool, machine learning (ML) has been demonstrated to simplify several cumbersome steps in data usage and interpretation in material science research, whilst also avoiding errors to a certain extent [12, 13]. ML utilises computing power to improve the generalisation ability and efficiency in materials science [14]. Today, ML is increasingly used in many aspects of material science, such as prediction of materials, material optimisation, crystal structure prediction, and material property prediction [15-20].

The traditional (empirical) production of alloy materials can be an expensive and difficult manner to approach the development of new materials [21]. Consequently, machine learning has an important role to play in helping inform engineers and scientists with respect to possible down-selection in alloy synthesis. There exists a desire for approaches, including high-throughput methods, relevant to the accelerated exploration of MPEAs [22, 23]. To date, based on the exploitation of experimental data, ML models have shown the ability to find relationships between alloy composition and alloy mechanical properties through model training. At the same time, the ML models may therefore predict the mechanical properties of alloy materials on demand [24]. Some examples from the domain includes the recent work of [25] who used linear regression (LR), gradient boosted regression (GBR), and random forest regression (RFR) models to analyse MPEAs. Such work was able to successfully predict the elasticity in composite multi-principal element alloys (MPEAs) by training the density functional theory (DFT) data set constant.

The present paper is not intended to be a comprehensive or exhaustive treatise of the topic of MPEAs or MPEA design. Instead, this work is the concise compilation of a study that explored the use of ML models upon a recently compiled dataset of MPEA properties – which the authors believe warrants dissemination as it may be useful for other researchers applying ML tools to metallic alloys, and also to provide some technical background to correspond with the availability of user web tool that readers may interact with (and hence, wish to know the key workings of the tool which they are utilising). The dataset used in the present work includes 1014 unique entries. The dataset employed has since been expanded and published [26] – making it freely available via open access (where the final dataset includes 1713 unique entries). The characteristics of the data used in the present work are included in Table 1, below.

**Table 1**. *Dataset characteristics for the ML model explored in the present study.*

| Mechanical Properties | Labeled data size | Unlabeled data size |
|---|---|---|
| Elongation | 179 | 835 |
| Hardness | 608 | 406 |
| Plasticity | 421 | 593 |
| Tensile strength | 195 | 819 |
| Yield strength | 256 | 758 |
| Compressive strength | 490 | 524 |

The overall workflow of the study herein is depicted in Figure 1. The workflow provides the key steps relevant to the work, which included the curation of the dataset (which was a significant manual human task), the application of an important parser tool to the curated data as described herein (further below), the broad exploration of ML models, and ultimately the selection of a model that forms the basis for a user tool.

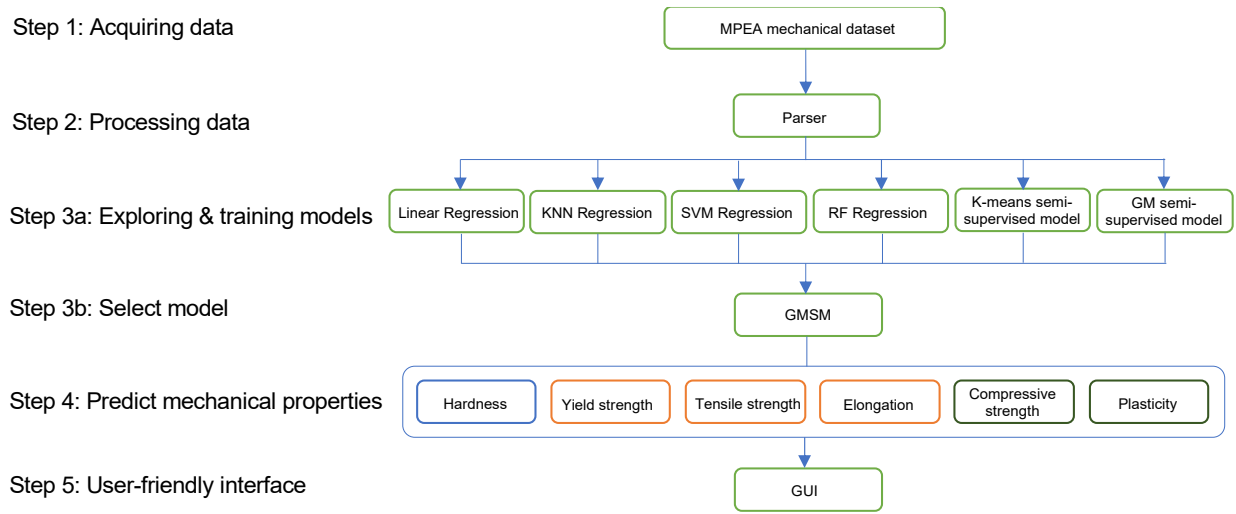

**Figure 1**. *The key steps and workflow in the present study.*

The present work, whilst primitive and considered an early-stage model, is the only user interactive MPEA tool capable of property prediction that is available freely, with some basic utility in helping inform MPEA design. The final model from the present work has known shortcomings, which include: not integrating MPEA processing/production method, and the absence of detailed feature analysis in mechanistic interpretations. As a result of this, subsequent work is underway from the authors that provides a deeper technical insight into model performance and feature analysis, using the largest known dataset [26] and incorporating a wider range of inputs (including alloy processing and calculated parameters).

## 2. Approach

### 2.1 Data and parser

The establishment of the MPEA dataset was primarily a process of manual collection from the literature. It took significant human effort to extract such information from the literature, which was a combination of evaluating tabulated data, or manually assessing plotted data.

The purpose of the present study is to predict the mechanical properties of MPEAs, from the composition (and elemental ratios) comprising an alloy. In order to facilitate the ML model to identify and analyse input content, this study has created and employed a "Parser" to process the data from the MPEA dataset. The Parser is used to process text (or, text + numerical values), in order to provide input data that is suitable for ML models. An example is seen in Figure 2, which shows the ability of the Parser tool to convert CoCrFeNi (which is an MPEA with constituent elements present in equiatomic proportions).

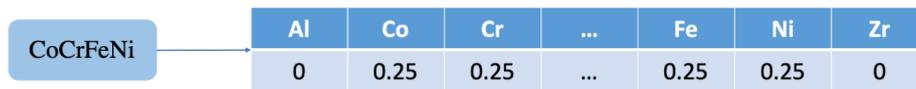

**Figure 2**. *Schematic representation of the Parser tool function, which is to convert an alloy description into structured data.*

The Parser plays a key role in alloy data processing; convert unstructured data into a structured format that facilitates data analysis. The parser script employed can be accessed at the GitHub repository: https://github.com/anucecszl/mechanical_property_parsing. The principle of the Parser can be as follows:

- Firstly, the Parser reads the chemical formula of an alloy (which is the first column in the mechanical property dataset).
- Secondly, it cleans redundant symbols in the chemical formula (such as spaces and brackets).
- Thereafter, it will normalise the proportion of different elements in the alloy composition. It will then provide the ratio for corresponding element and output a CSV file. The CSV file contains the 27 elements that appear in the dataset, along with the proportion of each element for each MPEA.
- The final output is utilised in machine learning processes.

For example, if the alloy composition is TiNbMoMnFe, the output of Parser would be $Ti_{0.2}Nb_{0.2}Mo_{0.2}Mn_{0.2}Fe_{0.2}$. The alloy chemical composition in the data set is converted into a form with a composition ratio of 1; examples of which may be seen in Figure 3.

|   | Al       | B   | C   | Co       | Cr       | ... | W   | Y   | Zn  | Zr  |
|---|----------|-----|-----|----------|----------|-----|-----|-----|-----|-----|
| 0 | 0.034188 | 0.0 | 0.0 | 0.256410 | 0.170940 | ... | 0.0 | 0.0 | 0.0 | 0.0 |
| 1 | 0.032520 | 0.0 | 0.0 | 0.243902 | 0.162602 | ... | 0.0 | 0.0 | 0.0 | 0.0 |
| 2 | 0.031008 | 0.0 | 0.0 | 0.232558 | 0.155039 | ... | 0.0 | 0.0 | 0.0 | 0.0 |
| 3 | 0.000000 | 0.0 | 0.0 | 0.333333 | 0.000000 | ... | 0.0 | 0.0 | 0.0 | 0.0 |
| 4 | 0.000000 | 0.0 | 0.0 | 0.333333 | 0.000000 | ... | 0.0 | 0.0 | 0.0 | 0.0 |

**Figure 3**. *Screen shot of a portion of the Parser output that reveals alloy name converted to individual alloy proportions allowing machine learning.*

The elements used (the Input parameters) for the ML models herein are listed in Figure 4, along with the properties that correlate with each alloy (the Output properties).

| Input Parameters | | | | | | | | Output Properties | |
|---|---|---|---|---|---|---|---|---|---|
| Al | B | C | Co | Cr | Cu | Fe | Ga | Ge | Hardness(HV) | Yield strength(Mpa) |
| Hf | Li | Mg | Mn | Mo | N | Nb | Ni | Sc | Tensile strength(Mpa) | Elongation(%) |
| Si | Sn | Ta | Ti | V | W | Y | Zn | Zr | Compressive strength(Mpa) | Plasticity(%) |

**Figure 4**. *Input parameters and Output properties for the ML modelling in this study.* .

Datasets can (inevitably) contain data errors and data set size limitations; along with the fact that data in the open literature also has incomplete 'Output properties' for each alloy. This is understandable, as not all studies in the literature seek to determine all properties, for each alloy produced and tested. In the database used for the study herein, a certain proportion of the six 'Output properties' are missing. Therefore, in order to address the issue of missing data this n the process of machine learning, an approach known as "data imputation" may be used. Data imputation is a technique for dealing with missing data, which can improve the completeness of data as described in a study by Zhang [27]. The application of data imputation (or not) rests with the respective ML model user, as to whether it is warranted or meaningful. When applied and beneficial, through data enhancement from data imputation, the so-called robustness of an ML model may be improved; such that can better adapt to new data. There are numerous methods of data enhancement, listed below:

1. Discarding / disregarding missing values (i.e. dropping null values).
2. Assuming / assigning null as 0.
3. Representing null by the mean value.
4. Filling nulls by interpolation.
5. Replacing nulls by algorithm fitting.

## 2.2. Models employed

In the present study, a number of ML models were explored, to allow user discretion as to the selection of a suitable ML model to be applied. The models explored are briefly described herein – including a concise implementation note for application of the models via python.

Supervised Models

When using supervised models, in order to reduce the impact of missing data on the prediction accuracy of the model, the nulls in the mechanical dataset were dropped. From the remaining data, 7% of the data was withheld and randomly selected as the test set.

- ***Linear Regression***: The first ML model chosen in this study is known as Ordinary Least Squares (OLS), which is simply a basic regression analysis model. The concept of OLS is to read the input data, assume that there is a specific linear relationship between the input and the output, and minimise the residual sum of squares to estimate the regression coefficient and intercept [28]. One or more best-fit lines (or curves) are thence selected to predict the continuous target variable such that the prediction error is minimised. The linear regression model is implemented by calling the OLS function in the python statsmodels library – and using the training set to fit the Linear Regression Model.
- **Support Vector Regression**: Support Vector Regression (SVR) regression [29] as the name implies, is based on regression analysis and classification [30]. By extracting and selecting features from samples in the data set, the low-dimensional space is mapped to the high-dimensional space through the kernel function so that the samples can be classified in the high-dimensional area. Then the hyperplane is determined by minimising the loss to predict the six mechanical properties (the Output properties). SVR calls an existing function in the python *sklearn* library. Through the function GridSearchCV () , a grid search on the parameters of SVR is conducted, and iterated 10 times. Then GridSearchCV () realises the optimisation of the model.
- **K-Nearest Neighbors Regression**: The K-Nearest Neighbors (KNN) regression model is nowadays a ubiquitous model for dealing with problems that require regression [31].

The approach for KNN regression is to assign the k samples that are closer to the target sample to one class through a weighted average in the eigenvalue space. The KNN regression model sorts the categories near unknown data according to a weight, and then determines the so-called category of the target predicted value. However, due to the missing values that are characteristic of the dataset used herein, the KNN regression algorithm is susceptible to leading to high classification error rates. The KNN Regression function may be implemented from the python *sklearn* library. Its optimisation method is as same as that of SVR and it is also implemented through *GridSearchCV* () with 10 iterations.

- **Random Forest Regression**: Random Forest Regression (RFR) predicts the output value (of a sample) by constructing multiple decision trees. When dealing with regression problems, RFR may improve model performance and the prediction result accuracy by adjusting hyperparameters [32] – given that the decision tree structure can be tuned by hyperparameters. In predicting an MPEAs mechanical properties, each alloy has 27 elements as inputs (inclusive of zero values for elements that are not in a given alloy). When the decision tree is split, m attributes are randomly selected, and hence another attribute is selected as the so-called split attribute. The random forest (RF) results are obtained from the results of a large number of decision trees. The RF Regression function can be applied from the python *sklearn* library. In order to achieve optimisation of RF model, the present work iterates the 'max depth' and 'number of estimator' within 100, in order to obtain the highest combination of high-accuracy parameters.

Semi-supervised Models

Semi-supervised model may be used to solve regression problems where there is a significant proportion of missing values in the data [33]. The basic principle of semi-supervised models is to use the labelled data in the original dataset to train an unsupervised model, then to use the predicted unlabelled data results as pseudo-labels, then to reuse the training data, and thence continue to iterate until the model converges. Thereafter, the use of a supervised model to predict mechanical properties can be applied.

Using a semi-supervised model may be done if one seeks to address the issue of missing values in the dataset, and the impact that may then have on the model itself. The working principle of a semi-supervised model is to shows that a trained model has a good generalisation ability, and therefore predicted results of the model will subsequently be more accurate. Using the basic idea of semi-supervised learning, we therefore first use an unsupervised learning model to 'fill' any unlabelled (i.e. missing) data; and then use a supervised learning model to train the data. The two such approaches explored in this study are outlined below.

- **K-means Clustering Semi-supervised Model**: The principle of the K-Means algorithm is to select k points as the cluster centres randomly. Each piece of data is assigned to the nearest centre, thus forming k clusters. Then, the approach recalculates the average value of each cluster to obtain a new centre point - iterating the above steps until the position of the centre point does not change [34]. Thereafter, we applied the Random Forest Regression (RFR) model to predict the mechanical properties.

  The supervised learning model (RFR) was applied without deleting any data of the original Parser results and taking 9% of the data from the non-null portion of the dataset as the test set for the RF regression model. This test set does not participate in any unsupervised learning process. Then we used K-Means to process the remainder of the non-null and all null data. The K-Means model uses existing function KMeans() in the python *sklearn.cluster* library. We set the cluster number in the K-Means model as 10, and iterated the model 100 times. In the unsupervised learning process, the average of the target mechanical properties is regarded as the label to fill the nulls. This method was carried out to improve the generalisation ability of the model. When fine using the RFR to predict mechanical properties, the parameter tuning for as was the same as that in the supervised model.

- **Gaussian Mixture Semi-supervised Model**: The Gaussian Mixture Semi-supervised Model (GMSM) is a probabilistic model. The GMSM first calculates the response of all data to each expectation-maximisation (EM) algorithm EM [35]. Then, the approach is to calculate the parameters of each sub-model based on the responsivity, before finally iterating to achieve the final result [36]. Compared with K-Means, GMSM calculates the parameters of each Gaussian distribution. At the same time, the GMSM realises the distribution parameters' solution by calculating the maximum value of the likelihood function. The GMSM method is the similar to that of the KMSM, except that the GMSM uses the function *GaussianMixture*() in the python *sklearn.mixture* library.

## 3. Results and discussion

The predictions from the supervised models for the prediction of the six mechanical properties were compared, and the fittest model (i.e. best performing) along with the evaluation metrics of the prediction results are shown in Table 2. When seeking to solve practical problems, the use of only a single evaluation criterion cannot comprehensively measure the performance of a model. In an attempt to more holistically measure the error between the predicted value and the actual value of the ML models, this study used two evaluation metrics, namely: Mean Absolute Error (MAE) and Coefficient of Determination ($R^2$) – which are represented by Eqn. 1 and Eqn. 2.

Table 2. *ML model performance for supervised models explored in the present study.*

| Mechanical Properties | Fittest Model | MAE ↓ | $R^2$ Score(%) ↑ |
|---|---|---|---|
| Elongation | RFR | 7.76 | 77.05 |
| Hardness | KNNR | 124.55 | 63.48 |
| Plasticity | RFR | 10.34 | 44.58 |
| Tensile strength | RFR | 140.85 | 83.16 |
| Yield strength | KNNR | 346.88 | 51.86 |
| Compressive strength | LR | 18.68 | 23.37 |

$$MAE = \frac{1}{n}\sum_{i=1}^{n}|(y_i - \hat{y}_i)|$$

............(Eqn1)

$$R^2 = \frac{\sum_{i=1}^{n}(y_i - \hat{y}_i)^2}{\sum_{i=1}^{n}(y_i - \overline{y_i})^2}$$

............(Eqn2)

In contrast to other evaluation criteria, the MAE is not sensitive to outliers. This is because the 'square' operation is not used in the determination of MAE, and hence the metric will not be significantly affected by a population of errors/outliers. For example, the compressive strength value in the data set varies greatly, however the MAE as an evaluation metric provides an insight that the range of the target variable is relatively large. Meanwhile, the coefficient of determination, is an evaluation index used to measure the degree of fit of the model.

The performance of the Semi-supervised models is provided via evaluation metrics provided in Tables 3a and 3b, below.

| Mechanical Properties | Evaluation Metrics | | Mechanical Properties | Evaluation Metrics | |
|---|---|---|---|---|---|
| | MAE ↓ | $R^2$ Score(%) ↑ | | MAE ↓ | $R^2$ Score(%) ↑ |
| Elongation | 8.26 | 82.26 | Elongation | 8.07 | 83.40 |
| Hardness | 118.34 | 66.86 | Hardness | 117.63 | 67.87 |
| Plasticity | 16.01 | 17.53 | Plasticity | 17.36 | 9.17 |
| Tensile strength | 204.22 | 77.51 | Tensile strength | 213.22 | 75.54 |
| Yield strength | 456.70 | 25.86 | Yield strength | 449.42 | 28.33 |
| Compressive strength | 568.22 | 27.68 | Compressive strength | 559.06 | 28.53 |

**Table 3**. *ML model performance for Semi-supervised models explored in the present study (a) KMSM, and (b) GMSM.*

According to the comparative assessment of the ML models performance summarised in Table 2 and Table 3, it can be surmised that the that the overall prediction of the RFR is the 'best' from the single regression models. It is also noted, that the overall prediction accuracies are not what would be deemed aspiration for the prediction of some of the mechanical properties (namely compressive strength and plasticity, which are derived from compression testing).

In evaluation of the prediction accuracy of the semi-supervised models, it was determined that the GMSM was the 'best', in a relative sense. In order to select a (single) ML model that could be applied as a user tool for the prediction of MPEA properties, an assessment between the RFR (single supervised regression models) and the GMSM model was also carried out by means of comparing the scatterplots of prediction results. The results are presented in Figure 5, from which it can be observed that the prediction result of GMSM conform to the true data in a (slightly) better manner than the RFR model. This visual analysis is important, because it supplements the results from the evaluation metrics alone. Furthermore, the visual analysis also rationalises that whilst the evaluation metrics provide values that are far from aspirational, the overall trends in the performance of a very primitive ML model are – to an appropriate extent – meaningful. From Figure 5, both models tend to either predict or underpredict the Output parameter values at what are the higher range of values (which is conservative), and also provide reasonable values for key parameters such as tensile strength.

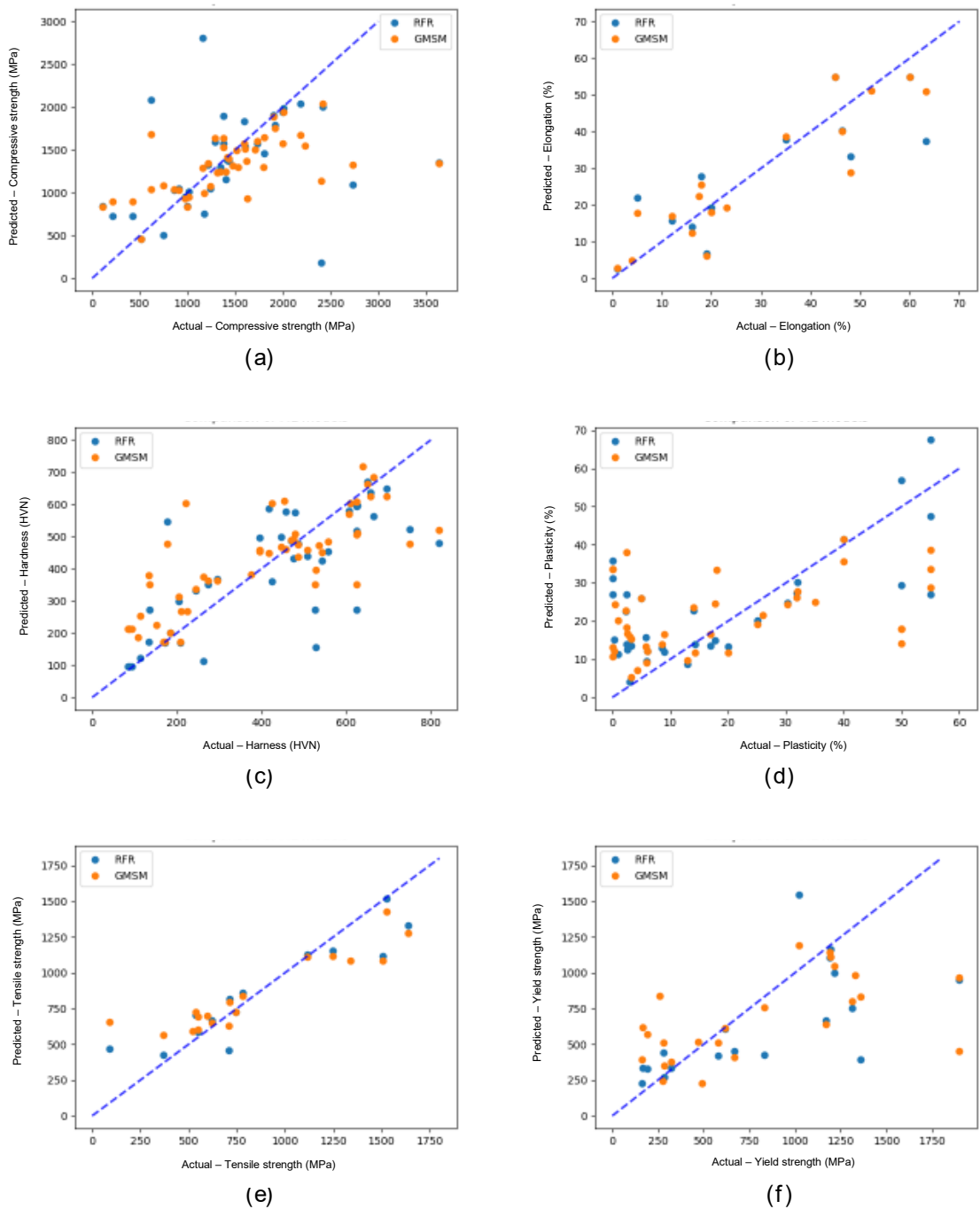

**Figure 5**. *Performance of the RFR and GMSM models, presented as actual vs. predicted, allowing a visual comparative assessment of ML model performance.*

Whilst the work herein has not conducted a thorough mechanistic analysis to correlate model performance with MPEA characteristics, some discussion points can be raised nonetheless. The prediction ability of the Random Forest Regression (RFR) ML appears to have been (comparatively) the best, pointing to the notion that there is likely a complex nonlinear relationship between the elemental composition of the alloy and the mechanical properties of the alloy. Based on this observation from the application of supervised models, proposing a random forest regression method using an unsupervised model to populate the data – was selected. As one exampled (from Figure 5) is that when predicting elongation, the prediction results from the GMSM were comparatively better than the RFR alone. The results herein have shown (albeit within the confines of the study, that did not include alloy processing as an Input parameter) that the introduction of the unsupervised model improved the quality of the data and thence the predictive performance of the RFR. Whilst not wholly conclusive, it can be posited that the semi-supervised learning model showed (some) good generalisation ability in dealing with the problem of low data labelling rate; for example, when predicting Hardness and Tensile Strength, the accuracy rates are reasonably high.

Compared with the several mechanical properties described above, the prediction accuracy of compressive strength from supervised models was the lowest. There may be several reasons for this result, and more broadly the results herein. Issues that are not elaborated but worth discussion (for future consideration) include:

- Inaccurate feature selection. For example, the compressive strength of an alloy may not only be related to elemental composition, but also be affected by other factors, such as processing, phases present, precipitate or intermetallic presence, etc. This is already acknowledged and known, because the database includes alloys of similar composition and different processing conditions with different mechanical properties.
- Model selection requiring more accurate feature extraction. For example, by spending more time with model tuning, better performance may be expected. From the domain of single principal element alloys, there is evidence that using neural networks and adjusting the parameters of the network, the accuracy of predicting compressive strength can be improved to a certain extent. In the research of [37], the researchers used the neural network to predict the compressive strength of Mg-Al-Zn alloys, and the accuracy rate reached 94%.
- Data size. The prediction results of Plasticity were the lowest among the models tested. The possible reasons are that the population of unlabelled dataset size was proportionally too large. The Output parameter of plasticity had among the lowest number of datapoints (with compressive strength), and also revealed it was not able to fit regression models well.
- The work herein does not harness mechanistic aspects of MPEA physical metallurgy, nor does it seek model interpretability. It is purely a data science approach and hence a primitive model. An aspect that will be difficult to manage (or more specifically, extract from the available data) is the quality and reproducibility of the experiments that went into the dataset. One issue is the mechanical property evaluation is often dependent on test protocols. Variations in test protocol may include the load in hardness testing, or the possibility of cross-sectional area reduction (or not) in tensile testing (and elongation evaluations), which will add an inherent variability and scatter into the dataset.

## 4. GUI

A Graphical User Interface (GUI) can make it simple for a user (which may be a non-expert, alloy designer, or simply a researcher that is unfamiliar with coding (python) or from a different field), to utilise the pre-trained model developed herein. The GUI developed for the present work, uses Gradio to realise a functional GUI. Gradio is a Python library that can be used to implement an interactive interface to machine learning models [38]. To make the GUI as widely accessible as possible, it was deployed web-based tool, that employs the Gaussian Mixture Semi-supervised Model (GMSM).

A key feature of the GUI, is it allows user interaction with the GMSM model, allowing the prediction of mechanical properties for MPEAs, based on a user defined MPEA composition (that may be a completely novel or unique composition) – with benefits for new alloy designs. New MPEA designs can be explored, which may include exploration of specific elemental combinations, or elements of desired density, cost, or availability, etc. To be most user friendly, the basic requirements of the GUI were a concise interface, comprehensive information, and simple operation. For example, using the Gradio GUI, users only need to input the MPEA alloy composition in the web interface to complete model testing. Through this method, the interactivity of the machine learning model can be improved,  The GUI is accessed at: https://huggingface.co/spaces/RuijiaTan/MultiPrincipalElementAlloyPropertyPredictor.

Some details of the user interface are summarised as follows:

- The Gradio GUI web page provides the name of the model and a brief description of the instructions for use (Section i, Figure 6).
- Then user enters the proportional values for each element they desire in their MPEA (Section ii, Figure 6), and then clicks "Predict"
- Then GUI will then provide the user their MPEA composition as an alloy ratio (Section iii, Figure 6) enters
- Then model will also then populate all of the outputs that are seen in 'Section iv' of Figure 6.

**Figure 6**. *The Gradio GUI web interface for the user-interactive deployment of the GMSM.*

The first 6 output parameters shown in Section iv of Figure 6, are the predicted mechanical properties (i.e. the mechanical properties of the use-input MPEA composition). The remaining outputs are the so-called Empirical Parameters, which were calculated automatically using the Empirical Properties Calculator, created by [23]; that calculates 12 empirical parameters. The empirical parameters calculated are described (along with their corresponding formula) in Table 4, and based on the determination approaches from [39-42]. In addition, the 12 empirical parameters, the alloy density is calculated (from the constituent element densities), and the

MPEA cost is estimated drawing from the market price of elements from the website created by [43].

**Table 4**. *The 12 empirical parameters automatically calculated in the Gradio GUI.*

| Empirical Parameter | Formula |
|---|---|
| Entropy of Mixing ($J/K * mol$) | $\Delta H_{mix} = \sum_{i=1, i \neq j}^{n} 4H_{i,j} c_i c_j$ |
| Average Atomic Radius ($Angstroms$) | $a = \sum_{i=1}^{n} c_i r_i$ |
| Atomic Size Difference | $\delta = 100 \times \sqrt{\sum_{i=1}^{n} c_i (1 - r_i/a)^2}$ |
| Enthalpy of Mixing ($kJ/mol$) | $\Delta S_{mix} = -R \sum_{i=1}^{n} c_i ln(c_i)$ |
| Standard Deviation of Enthalpy | $\sigma_{\Delta H} = \sqrt{\sum_{i=1, i \neq j}^{n} c_i c_j (H_{i,j} - \Delta H_{mix})^2}$ |
| Average Melting Point ($Tm, in Celcius$) | $T_m = \sum_{i=1}^{n} c_i T_i$ |
| Standard Deviation of Melting Point | $\sigma_{T_m} = 100 \times \sqrt{\sum_{i=1}^{n} c_i (1 - T_i/T_m)^2}$ |
| Average Electronegativity ($X$) | $X = \sum_{i=1}^{n} c_i X_i$ |
| Standard Deviation of Electronegativity | $\Delta X = 100 \times \sqrt{\sum_{i=1}^{n} c_i (1 - X_i/X)^2}$ |
| Valence Electron Concentration ($VEC$) | $VEC = \sum_{i=1}^{n} c_i VEC_i$ |
| Standard Deviation of Valence Electron Concentration ($VEC$) | $\sigma_{VEC} = \sqrt{\sum_{i=1}^{n} c_i (VEC_i - VEC)^2}$ |
| The Unitless Parameter Omega | $\Omega = T_m \Delta S_{mix} / |\Delta H_{mix}|$ |

$c_i$: The molar ratio of the $i_{th}$ element in the alloy composition.
$r_i$: The atomic radius of the $i_{th}$ element in the alloy composition.
$R$: The ideal gas constant.

## 5. Future work

Although this project demonstrated a workflow to (i) train an ML model on the MPEA database, and (ii) predict the mechanical properties of MPEAs (including any user input composition of MPEA); there are still many aspects of future work that would improve the accuracy of predictions. For transparency, the authors also acknowledge that the final model presented herein is not a so-called field deployable tool that can be used with high confidence. It does however, reveal a number of trends, and approximate values in most cases, and the associated automated calculation of empirical parameters is also very useful.

A number of considerations that are beneficial in future iterations of this (and similar) models may include – at a minimum:

- More training data. In the process of data collection, it was evident that there were some errors and omissions in many works. In addition, many papers only assess a small number of Output properties. It is acknowledged that the dataset size employed herein, although field leading at the time, is considered (by ML standards) a very small dataset.

- Minimising systemic errors. For example, in the extraction of data from graphical sources, there is manual reading and selection. Access to numerical data would be more significant. Also, all Output properties are considered to be similar in the present study; however, if testing included methods for strength determination that included cross-sectional area reduction (or not), the consistency of results is an area that is un-explored.

- This study investigated applying ML models to predict various properties of MPEAs based on the alloy composition. However, there are a range of empirical parameters, such as the enthalpy of mixing, valence electron concentration, and average bulk modulus, that have been demonstrated to correlate with the properties of MPEAs [42, 44]. Future work could consider incorporating these parameters into the modelling to further enhance the accuracy and granularity of predictions. It's worth noting that adding too many features may lead to a model that's overly complex and at risk of overfitting. Therefore, performing feature evaluation to avoid overfitting will be a crucial step.

- The processing conditions of the MPEAs should be considered as part of input of models. Through the analysis of the data, it can be seen that the mechanical properties of the alloys with the same composition are different – and this is a result of different production and processing conditions. The authors expect that the inclusion of MPEA production and processing as an input (in addition to alloy composition) will improve the prediction performance (albeit to a yet unknown) extent.

- Whilst the present study explored a number of ML models, the selection of the models and the adjustment of the parameters associated with such ML models can also significantly affect the model's prediction results. The model selection deserves critical attention. Furthermore, considering that generative adversarial networks (GANs) can generate realistic (but synthetic) sample data, the exploration of GANs may also lead to models that present more reliable results.

## 6. Conclusions

Overall, this work revealed a workflow for machine learning (ML) to be applied to MPEAs, from data processing, to a user tool. Bearing in mind the stated limitations of the approach used herein, the user tool (that operates via a web-based GUI) is a compelling case for justification of ML approaches that can be harnessed for improving the efficiency of synthesising MPEAs and predicting the mechanical properties of such alloys.

In this project, we first created a database / dataset, which included alloy composition and six mechanical properties (Output properties). A Parser tool was developed and applied to the MPEAs in order to parse each alloy composition in the dataset into the form of an input vector for ML models. The prediction results from a number of models were explored and compared, including Linear Regression, K-Nearest Neighbor Regression (KNNR), Support Vector Regression (SVR), and Random Forest Regression (RFR) – upon the mechanical properties of MPEAs. Additionally, using K-Means clustering and GMM to realise data imputation – accuracy was improved through Random Forest Regression – although this was noted as not being a general approach for future work (although it was the so-called best approach for the study herein). The study is summarised by a web-based GUI that was implemented through Gradio. The GUI includes a handy empirical calculator that can be used to calculate the corresponding 14 empirical parameters of any MPEA that a user may conceive – which also adds utility to the field of research and education regarding MPEAs.

**Acknowledgements**

This work was performed as the Research Project in COMP8604 by Ms. Ruijia Tan. We acknowledge the technical assistance of following individuals in the curation (including data mining) of the database presented herein: Theo Darmawan, Ninad Bhat, Himadri Shekhar Mondal and Dr. Zhuoran Zeng.

**Author contributions**

RT performed the coding. RT and ZL developed the study framework. RT contributed to manuscript preparation, and NB and ZL designed the study and also contributed to manuscript preparation. SZ and ZL also provided technical assistance. NB supervised the study.

**Competing interests**

The authors declare no competing interests.